\documentclass[superscriptaddress,twocolumn,pre,amsmath,showpacs]{revtex4-1}
\usepackage{graphicx}
\usepackage{dcolumn}
\usepackage{bm}
\usepackage{colordvi}
\usepackage{epsfig}
\usepackage{graphicx}
\usepackage[usenames]{color}

\newcommand{\re}{\text{Re}}

\newcommand{\ra}{\big\rangle} 
\newcommand{\la}{\big\langle}

\begin{document}

\title{Active Microrheology of Driven Granular Particles}

\author{Ting Wang}
\affiliation{Institut f\"ur Materialphysik im Weltraum,  
Deutsches Zentrum f\"ur Luft- und Raumfahrt (DLR), 51170 K\"oln, Germany}
\author{Matthias Grob}
\affiliation{Georg-August-Universit\"at G\"ottingen,  
Institut f\"ur Theoretische Physik, Friedrich-Hund-Platz 1, 
37077 G\"ottingen, Germany} 
\author{Annette Zippelius} 
\affiliation{Georg-August-Universit\"at G\"ottingen,  
Institut f\"ur Theoretische Physik, Friedrich-Hund-Platz 1, 
37077 G\"ottingen, Germany} 
\affiliation{Max-Planck-Institut f\"ur Dynamik und Selbstorganisation, 
Am Fa\ss berg 17, 37077 G\"ottingen, Germany}
\author {Matthias Sperl}
\affiliation{Institut f\"ur Materialphysik im Weltraum,
Deutsches Zentrum f\"ur Luft- und Raumfahrt (DLR), 51170 K\"oln, Germany} 
 
\begin{abstract}

When pulling a particle in a driven granular fluid with constant force 
$F_{ex}$, the probe particle approaches a steady-state average velocity 
$v$. This velocity and the corresponding friction coefficient of the probe 
$\zeta=F_{ex}/v$ are obtained within a schematic model of mode-coupling 
theory and compared to results from event-driven simulations. For small 
and moderate drag forces, the model describes the simulation results 
successfully for both the linear as well as the nonlinear region: The 
linear response regime (constant friction) for small drag forces is 
followed by shear thinning (decreasing friction) for moderate forces. For 
large forces, the model demonstrates a subsequent increasing friction in 
qualitative agreement with the data. The square-root increase of the 
friction with force found in [Fiege et al., Granular Matter 
$\boldsymbol{14}$, 247 (2012)] is explained by a simple kinetic theory.

\end{abstract}

\date{\today}
\pacs{82.70.Dd, 64.70.Pf, 83.60.Df, 83.10.Gr}

\maketitle

\section{Introduction}

Active microrheology (AM) studies the mechanical response of a 
many-particle system on the microscopic level by pulling individual 
particles through the system either with constant force or at constant 
velocity \cite{Squires2010b}. While in passive microrheology only the 
linear response can be probed, AM can also be applied to explore the 
non-linear response by imposing large drag forces. An external force 
$F_{ex}$ can be imposed by magnetic \cite{Habdas2004} or optical tweezers 
\cite{Meyer2006} to a probe particle embedded in a soft material and then 
the responding steady-state velocity $\la v\ra$ is measured by optical 
microscopy \cite{Conchello2005}. Recently, AM experiments 
\cite{Habdas2004} and simulations \cite{Gnann2011b,Winter2012,Winter2013b} 
for dense colloidal suspensions found that (i) in the linear-response 
region, the friction coefficient of the probe $\zeta=F_{ex}/\la v\ra$ 
directly indicates the increasing rigidity of the system when approaching 
the glass transition from the liquid state; (ii) in the nonlinear response 
region, the friction coefficient tends to decrease to a certain value with 
increasing pulling force -- an effect reminiscent of shear thinning in 
macrorheology. Both effects could be explained by an extension of 
mode-coupling theory (MCT) to describe AM \cite{Gazuz2009,Gazuz2013}.

Within the MCT interpretation, the description of AM for colloidal 
suspensions is based on the existence of a glass transition in such 
systems. The interplay between growing density correlations by glass 
formation and the suppression of those correlations by microscopic shear 
explains the observed behavior of the friction microscopically. In 
addition to colloidal suspensions, a glass transition is also predicted by 
MCT for driven granular systems \cite{Kranz2010,Sperl2012, Kranz2013}. 
Here, the energy lost in the dissipative interparticle collisions is 
balanced by random agitation. Starting from the non-equilibrium steady 
state of this homogeneously driven granular system, the corresponding AM 
shall be elaborated below.

In AM of granular matter similar phenomena as in colloidal suspensions are 
found: (i) Dramatic increasing of the friction coefficient and (ii) shear 
thinning have been identified in experiments with horizontally vibrated 
granular particles \cite{Candelier2009a,Candelier2010b}, and both effects 
were reproduced in recent simulations of a two-dimensional granular system 
\cite{Fiege2012}. Moreover, for large pulling forces in the simulation, 
beyond the thinning regime the friction coefficients increase again and 
exhibit power-law behavior close to a square root: $\zeta(F_{ex}) \propto 
\sqrt{F_{ex}}$ for $F_{ex}\gg 1$. This finding is in contrast to the 
predicted constant friction (second linear regime) in the colloidal 
hard-sphere system \cite{Gazuz2009,Gazuz2013}. In the following, we shall 
demonstrate how a schematic MCT model can capture the increase with 
friction for large forces. In addition, for dilute systems we shall derive 
a square-root law for large forces exactly.

\section{Dynamics of a Granular Intruder}

The driven granular system is comprised of N identical particles 
interacting with each other. One probe particle experiences a constant 
pulling force $\bm{F}_{ex}$. The dynamics of the system is given for 
every particle $i$ by the equation of motion
\begin{equation}\label{eom}
     m\dot{\bm{v}}_{i}=-\zeta_{0} \bm{v}_{i}+\bm{f}_{int}^{i}
     +\bm{\eta}_{i}
     +\bm{F}_{ex}\delta_{i,s},
\end{equation}
where $\zeta_{0}$ is the bare friction depending on the friction of the 
surrounding medium, $\bm{f}_{int}^{i}$ is the particle interaction force, 
$\bm{\eta}_{i}$ is a random driving force satisfying a fluctuation 
dissipation relation $\la \bm{\eta}_{i}(t)\bm{\eta}_{j}(t')\ra = 
2\zeta_{0} k_{B}T/m\delta_{i,j}\delta(t-t')$, and the constant pulling 
force $\bm{F}_{ex}$ is imposed on the probe particle (denoted $s$) only.

\subsection{Schematic Model}   

The friction coefficient of the probe can be calculated by the 
integration-through-transition (ITT) method combined with the MCT 
approximation. This procedure was first applied to describe the 
macrorheology \cite{Fuchs2002} and was later extended to AM for colloidal 
suspension \cite{Gazuz2009,Gazuz2013}. We follow the approach in 
\cite{Gazuz2009,Gazuz2013} to construct a schematic MCT model for driven 
granular systems. Different from colloidal systems, the equations of 
motion for the density autocorrelation functions for both the bulk system 
and the probe particle, $\phi_{\bm{q}}^{s}(t):= \langle 
\rho_{\bm{q}}(t)\rho_{\bm{q}}^{*} \rangle/\langle \rho_{\bm{q}} 
\rho_{\bm{q}}^{*}\rangle$ and $\phi_{s,\,q}(t):=\langle \exp[ 
i\bm{q}\cdot(\bm{r}_{s}(t)-\bm{r}_{s})] \rangle$, respectively, include a 
second time derivative, because granular systems are not overdamped. Here, 
$\langle\cdots \rangle$ denotes the ensemble average and 
$\rho_{\bm{q}}(t):=\sum_{i=1}^{N}e^{i\bm{q} \cdot \bm{r}(t)}$ is the 
Fourier transform of the density. As usual for schematic models we ignore 
the dependence on the wave vector $\bm{q}$ of the correlation functions 
and model the memory kernel by a non-linear function of the correlation 
functions as follows:
\begin{equation}\label{phis} \begin{split} 
&\ddot{\phi}(t)+\nu\dot{\phi}(t)+\Omega^{2}\big[\phi(t)
                    +\int_{0}^{t}d\tau\,m(t-\tau)\dot{\phi}(\tau)\big]=0\\
&\ddot{\phi}_{s}(t)+\nu_{s}\dot{\phi}_{s}(t)+\Omega_{s}^{2}\big[\phi_{s}(t)
+\!\!\int_{0}^{t}\!\!d\tau\,m_{s}(t-\tau)\dot{\phi}_{s}(\tau)\big]=0\\
\end{split}
\end{equation}
with
\begin{equation}\label{mem}
\begin{split}
m(t)&=v_{1}\phi(t)+v_{2}\phi^{2}(t)\\
m_{s}(t)&=v_{A}\phi(t)\re[\phi_{s}(t)],
\end{split}
\end{equation}
where $m(t)$ and $m_{s}(t)$ are the memory kernels of the well-known 
$F_{12}$ model \cite{Goetze2009}. The control parameters of the host 
system are set to be $\nu=\Omega=1$, the host system is assumed to be 
large enough that its density correlation functions will not be affected 
by the external pulling force $F_{ex}$. The state points $(v_{1}, v_{2}) = 
\big(v_{1}^{c}(1+\sigma),\,v_{2}^{c}(1+\sigma)\big)$ are specified by a 
distance $\sigma$ to the transition line given by $v_{1}^{c} = 
v_2^{c}(2/\sqrt v_{2}^{c}-1)$ with the specific choice $v_{2}^{c} = 2$ for 
the transition point. $v_{A}$ indicates the coupling strength between the 
probe and the host system. $\Omega_{s}^{2}$ is the effective frequency of 
the correlation function of the probe and set to be 
$\Omega_{s}^{2}=1-iF_{ex}$. This can be obtained exactly from 
Eq.~(\ref{eom}) by considering the limit of vanishing interacting force:
\begin{equation}\label{eq:phislimit}
\phi_{\bm{q}}^{s}(t)= \exp(-\Omega_{s\,\bm{q}}^{2}t)
\,, \Omega_{s\,\bm{q}}^{2}= 
\frac{\bm{q}}{\zeta_{0}}\cdot(\bm{q}k_{B}T-i\bm{F}_{ex})\,.
\end{equation}
$\nu_{s}$ describes the dynamics of the density correlator of the probe 
for short time scales. In order to assure that $|\phi_{s}(t)|\leq 1$, it 
is required that
\begin{equation}\label{cvc} 
\nu_{s}>F_{ex}\,,
\end{equation} 
which is obtained by solving the second equation in Eqs.~(\ref{phis}) 
without the memory kernel. Note that the force-dependent $\nu_{s}(F_{ex})$ 
is different from the schematic model in equilibrium systems, in which 
$\nu_{s}$ is set to be constant \cite{Goetze2009}. The force dependence of 
$\nu_{s}$ indicates that the external pulling force affects the short time 
dynamics of the probe particle. By integration of the density 
autocorrelators \cite{Gazuz2009,Gazuz2013}, we get the expression for the 
effective friction of the probe as
\begin{equation}\label{eq:zeta}
 \zeta/\zeta_{0}=1+\int_{0}^{\infty}dt\,\text{Re}[\phi_{s}(t)]\phi(t)
\end{equation}

\begin{figure}[htbp]
\includegraphics[width=\columnwidth]{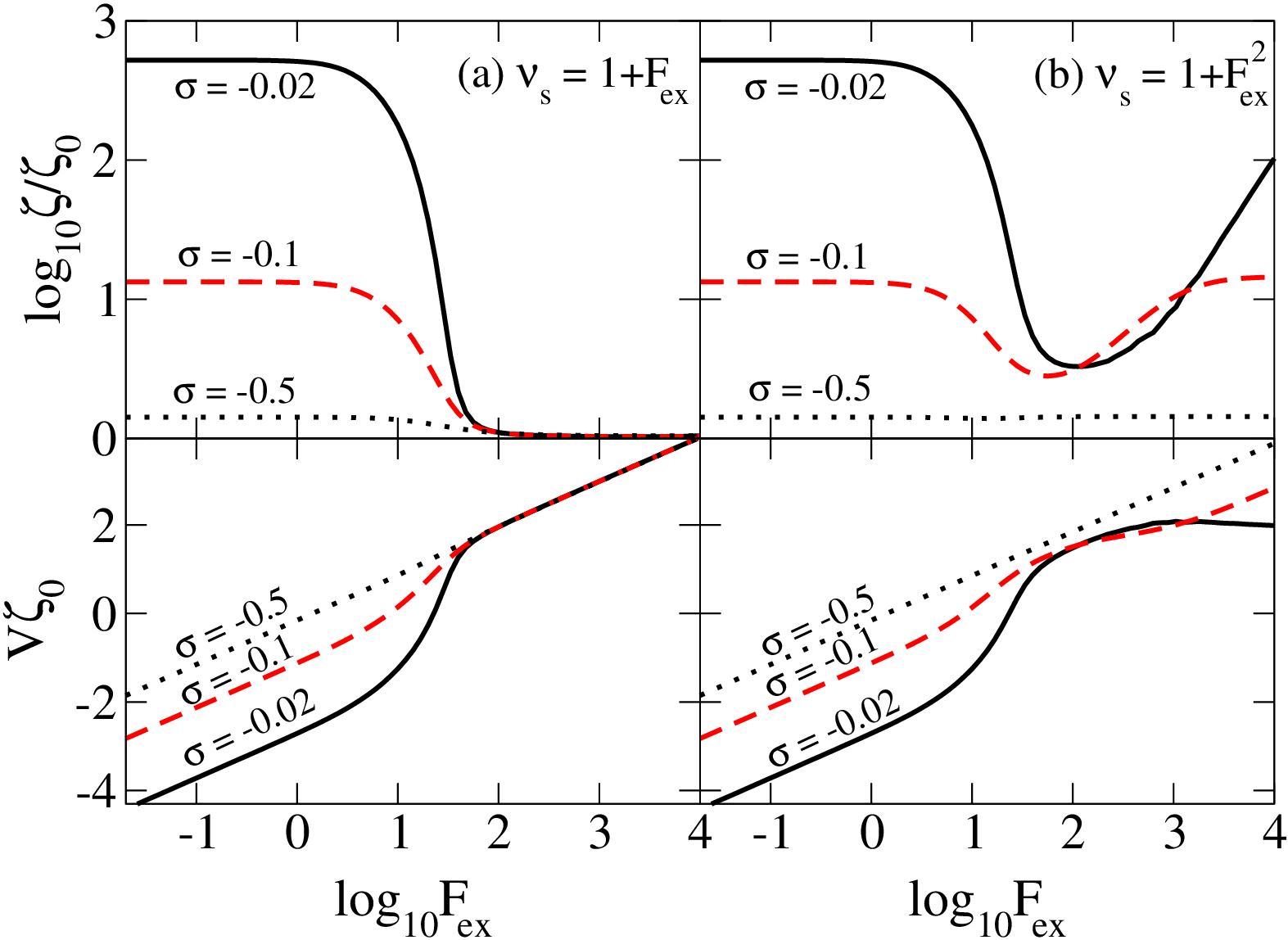}
\caption{\label{zf} Force-dependent friction coefficient (upper panels) 
for different models of the damping $\nu_{s}(F_{ex})$ together with the 
steady-state velocities (lower panels). $\sigma$ specifies the distance 
from the glass-transition point in the schematic model.}
\end{figure}

We propose two possible sets of $\nu_{s}(F_{ex})$ complying with the 
constraint in Eq.~(\ref{cvc}): (a) $\nu_{s}=1+F_{ex}$ and (b) 
$\nu_{s}=1+F_{ex}^{2}$. The respective numerical solutions for the 
force-dependent friction coefficients are given in Fig.~\ref{zf}, where 
$\sigma$ indicates the distance from the glass transition.

The force-dependent friction of the probe exhibits three characteristic 
regimes. For small pulling forces in both models, the friction coefficient 
is constant, or equivalently, the average velocity of the probe is 
proportional to the external pulling force. This region extends to 
external forces of order unity and describes a linear response. When 
approaching the glass transition, the friction increases drastically as 
the correlation functions in Eq.~(\ref{eq:zeta}) extend to increasingly 
longer time scales. Starting around $F_{ex}\approx 1$, the linear-response 
regime ends and gives rise to shear thinning: The friction decreases and 
it is proportionally easier to pull the particle. Equivalently, the 
average velocity of the intruder increases faster than linear with 
external force. For the model in the left panel of Fig.~\ref{zf}, the 
friction $\zeta$ approaches the limiting value given by the bare friction 
$\zeta_0$ and remains there for yet higher forces. This model hence 
describes behavior similar to the colloidal results for Newtonian 
microscopic dynamics. For the model in the right panel of Fig.~\ref{zf}, 
the friction approaches a minimum around $F_{ex}\approx 100$ and starts 
increasing for higher pulling forces.

In comparison, the models in Fig.~\ref{zf}(a) and Fig.~\ref{zf}(b) are 
almost equivalent in the linear-response and shear-thinning regimes, where 
the friction in Eq.~(\ref{eq:zeta}) is dominated by the memory effects 
leading to a slowing down of the relaxation. The difference in the 
microscopic damping $\nu_s$ does not play a significant role. In contrast, 
for large pulling forces, the correlation function $\phi_s(t)$ relaxes to 
zero rapidly and the integral~(\ref{eq:zeta}) is dominated by the 
short-time part of the correlation functions.

\subsection{Comparison with simulation data}

\begin{figure}[htbp]
\includegraphics[width=\columnwidth]{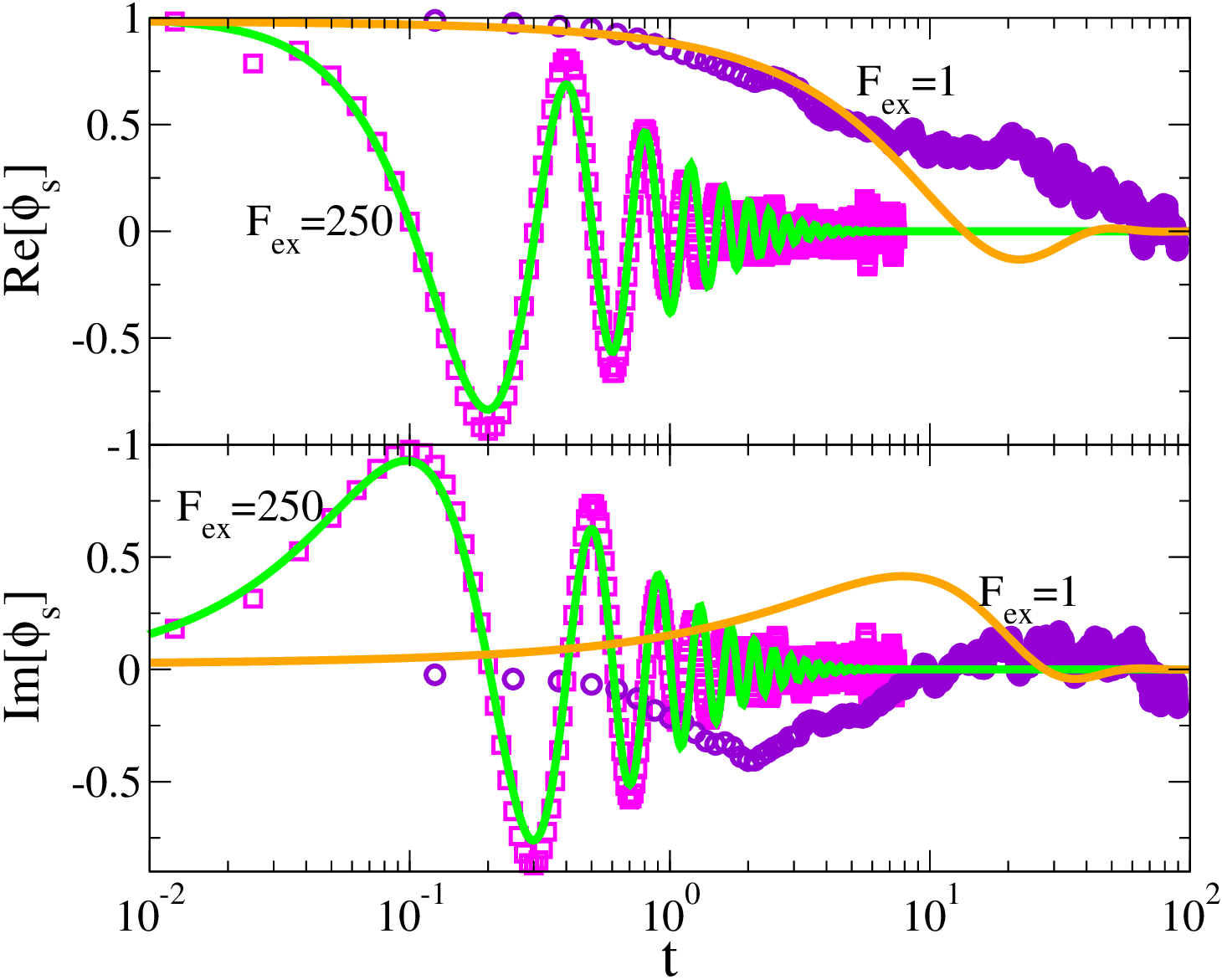}
\caption{\label{cor_fit} Comparison between the schematic model and the 
simulation data for the density autocorrelator of the probe 
particle at fixed packing fraction $\varphi=0.8$ and energy dissipation 
$\varepsilon=0.9$. The pulling forces are $F_{ex}=1$ and 250. The dashed 
lines represent the simulation data, the solid lines are the descriptions 
by the schematic model.}
\end{figure}

We adopt with $\nu_{s}=1+F_{ex}^{2}$ the second schematic model to compare 
with the simulation data in detail. The simulation setup is the same as 
described in \cite{Fiege2012}: in a bidisperse mixture of hard spheres 
with size ratio $R_s/R_b = 4/5$ of small to big particles and a respective 
mass ratio $m_s/m_b = 16/25$, an intruder of radius $R_0 = 2R_s$ and mass 
$m_0 = 4m_s$ is suspended. Lengths and masses are measured such that $R_s 
= 1$ and $m_s = 1$ and a time scale is set by requiring T=1 in the system 
with $F_ex = 0$, i.e. the random driving balances the dissipation by bare 
friction and collisions. Figure~\ref{cor_fit} shows the fit of the 
measured correlation functions by the model. The numerical solution of the 
density autocorrelator of the intruder fits quite well the corresponding 
simulation data for the moderately high force $F_{ex}=250$. For the 
smaller force $F_{ex}=1$, it shows some deviations. The fitting parameters 
are $v_{A}=200$ , $\sigma=-0.05$ for $\varepsilon = 0.9$ as well as (not 
shown in Fig.~\ref{cor_fit}) $v_{A} = 300$, $\sigma = -0.09$ for 
$\varepsilon=0.7$ and $v_{A}=600$, $\sigma = -0.13$ for $\varepsilon = 
0.1$. The other parameters are the same as the ones mentioned above.

\begin{figure}[htbp] 
\includegraphics[width=\columnwidth]{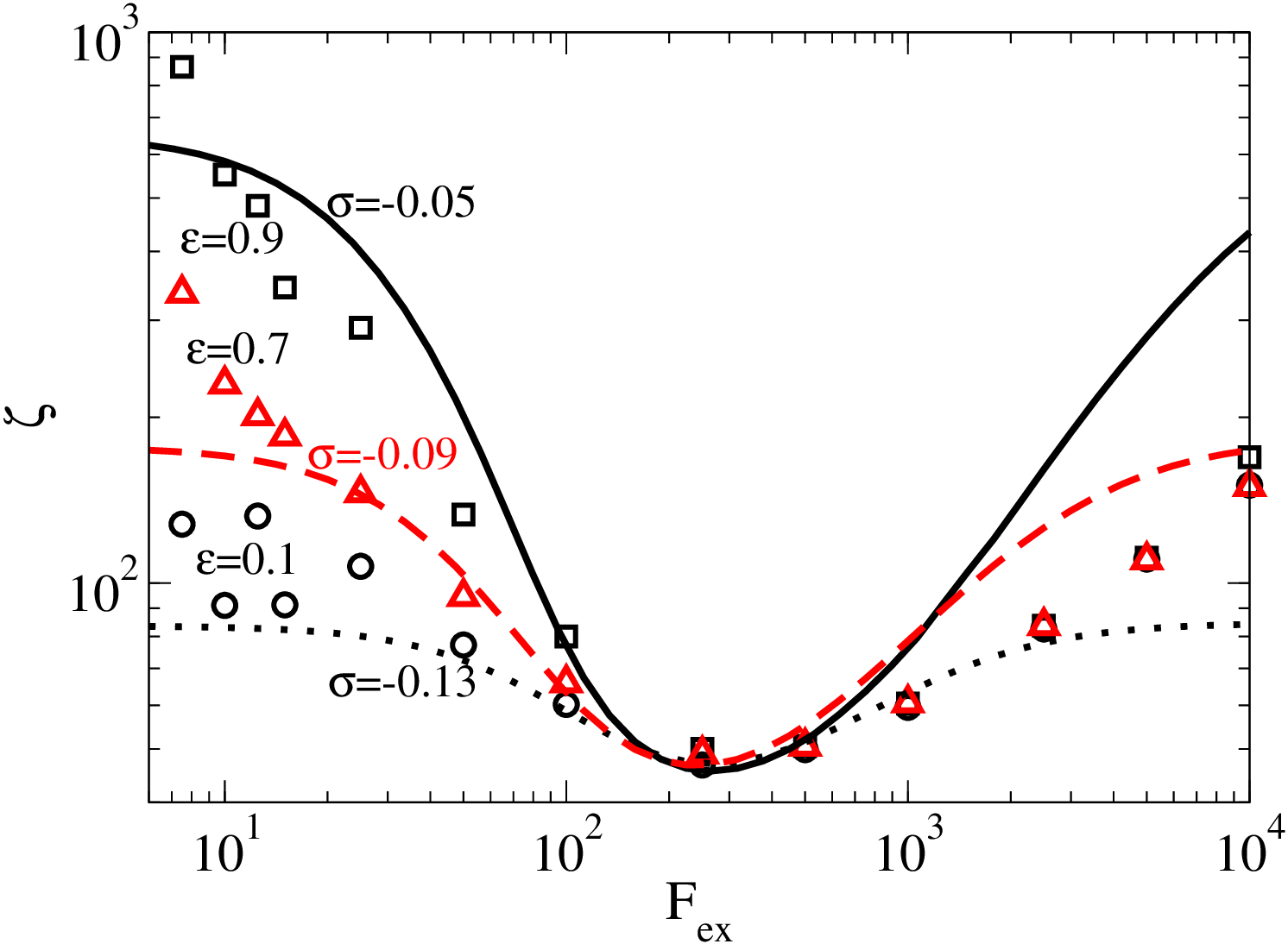} 
\caption{\label{fri_fit} Effective friction of the intruder for different 
energy dissipation $\varepsilon=0.9$, 0.7 , and 0.1 at fixed packing 
fraction $\varphi=0.8$ of the host fluid. Individual data points 
show the simulation results, curves represent the results from the 
schematic models.}
\end{figure}

The corresponding fit of the friction coefficients is given in 
Fig.~\ref{fri_fit}. In the regime of small forces, the schematic model 
shows a linear-response plateau. The simulation data also show a plateau 
for small forces (see Fig.~3 in \cite{Fiege2012}). As the glass transition 
is approached, this regime moves to smaller forces, so that it is visible 
in Fig.~\ref{fri_fit} only for the smallest $\varepsilon=0.1$ which is 
further away from the glass transition than $\varepsilon =0.9$ and $0.7$. 
However for $\varepsilon=0.1$ and $\varphi=0.8$ the simulations become 
increasingly difficult for small forces due to the occurrence of long 
lasting contacts. Hence the error bars become comparable to the result 
itself. For large pulling forces, the model shows qualitatively how the 
increasing friction coefficient can be rationalized within a schematic 
model. While the schematic model exhibits different limits for varying 
distances from the glass transition, the simulation data follow the same 
curve for $F_{ex} \gtrsim 500$. Between the extreme regimes of large and 
small pulling forces the friction coefficient exhibits a minimum that is 
similar for all distances from the glass transition for both the schematic 
model and the simulation.

While the schematic model may only qualitatively fit the simulation data 
for the friction coefficient in Fig.~\ref{fri_fit}, in addition to the 
good agreement of the correlation functions in Fig.~\ref{cor_fit} the 
results are also consistent with the predictions from MCT for hard spheres 
\cite{Kranz2010,Sperl2012,Kranz2013}. For smaller dissipation, i.e. larger 
coefficient of restitution $\varepsilon$ in Fig.~\ref{fri_fit}, the data 
can be described only by choosing points closer to the glass-transition 
line in the schematic model. Smaller distances to the glass transition 
given by smaller values of $\sigma$ indicate that for the same density of 
$\varphi = 0.8$ the data for $\varepsilon = 0.9$ are much closer to the 
glass transition than for $\varepsilon = 0.1$ with $\varepsilon = 0.7$ 
located in between. This finding is in agreement with the predicted 
increase of the glass-transition density with decreasing $\varepsilon$ 
within MCT \cite{Kranz2010,Sperl2012,Kranz2013}.

\section{Kinetic Theory in low density limit}

To clarify the origin of the scaling law $\zeta\propto \sqrt{F_{ex}} $ in 
the large-force asymptote, we propose a simple kinetic theory in the 
following. The simulation result from \cite{Fiege2012} has shown that the 
scaling law is independent of packing fraction. Therefore, a potential 
explanation of the increased friction by jamming or shear thickening 
seems unlikely. Also, the correlation functions for large pulling forces 
decay relatively quickly, cf. Fig.~\ref{cor_fit}, also contradicting a 
buildup of long-time glass-like contributions to the integrals like in 
Eq.~(\ref{eq:zeta}). In the following we shall therefore focus on the low 
density limit, where exact solutions can be obtained.

The formal solution of $\bm{v}(t)$ in Eq.~(\ref{eom}) can be readily 
obtained and the corresponding ensemble average of the velocity of the 
intruder is given by
 \begin{equation}\label{eq:sol}
    \la\bm{v}(t)\ra=\frac{\bm{F}_{ex}}{\zeta_{0}}\big(1-e^{-\frac{\zeta_{0}}{m}t}\big)+\frac{e^{-\frac{\zeta_{0}}{m}t}}{m}\int_{0}^{t}
    e^{\frac{\zeta_{0}}{m}t^\prime}\la\bm{f}_{int}(t^\prime)\ra dt^\prime
 \end{equation}
where we have averaged out the initial velocity and the random force: $\la 
\bm{v}_{0}\ra=0$ and $\la\bm{\eta}(t)\ra=0$.

The direct calculation of $\la\bm{f}_{int}(t)\ra $ in Eq.~(\ref{eq:sol}) 
is difficult. The key point of our kinetic theory is to introduce the mean 
free path of the intruder, $l_{0}=\rho^{-1}\sigma_{cr}$, where $\rho=N/V$ 
is the particle number density and $\sigma_{cr}$ is the intruder's cross 
section, which for hard sphere reduces to $\sigma_{cr}=4\pi R^{2}$. Let us 
denote the collision time as $t_{c}$. Between two successive collisions 
$nt_{c}<t<(n+1)t_{c}$, there is no interaction force in the hard sphere 
limit, $\la\bm{f}_{int}(t)\ra =0$. On average, after $t_{c}$ a collision 
event causes a momentum transfer from the intruder to its collision 
partner of the order of the intruder's complete momentum. The velocity of 
the intruder increases again from almost zero due to the constant pulling 
force. Statistically, the intruder's velocity exhibits periodic motion. 
Consider the motion of the probe in the first period: The average velocity 
reads
 \begin{equation}\label{av1}    
\la\bm{v}(t)\ra = \frac{\bm{F}_{ex}}{\zeta_{0}}
\big(1-e^{-\frac{\zeta_{0}}{m}t}\big),\quad 0\leq t\leq t_{c}\,,
 \end{equation}
and the displacement of the motion satisfies
\begin{equation}\label{key}
l_{0}=\Big|\int_{0}^{t_{c}}\la\bm{v}(t)\ra dt\Big|=
\frac{F_{ex}}{\zeta_{0}}\big[t_{c}-\frac{m}{\zeta_{0}}
(1-e^{-\frac{\zeta_{0}}{m}t_{c}})\big]\,.
\end{equation}
The average velocity of the probe is given by
\begin{equation}\label{key2}
\la v_{s}\ra=\frac{1}{t_{c}}\Big|\int_{0}^{t_{c}}\!\la\bm{v}(t)\ra dt\Big|
=\frac{l_{0}}{t_{c}}\,.
\end{equation}

In general the friction of the probe $\la v_{s}\ra$ can be calculated by 
Eqs.~(\ref{key},\ref{key2}) exactly. We first consider the two limiting 
cases $t_{c}\gg \frac{m}{\zeta_{0}}$ (overdamped limit) and $t_{c}\ll 
\frac{m}{\zeta_{0}}$ (ballistic regime).

In the overdamped limit, velocity relaxation dominates over collisions 
and the collision times are large, 
 \begin{equation}
 t_{c}=\frac{l_{0}\zeta_{0}}{F_{ex}}\gg\frac{m}{\zeta_{0}}
 \end{equation}
or equivalently,
\begin{equation}
\frac{F_{ex}}{\zeta_{0}^{2}}\ll\frac{l_{0}}{m}.
\end{equation}  
The average velocity and the friction of the intruder can be obtained by 
Eq.~(\ref{key2}) and definition of the friction itself, yielding
\begin{equation}
 \la v_{s}\ra =F_{ex}/\zeta_{0},\quad \zeta =\zeta_{0}\,.
\end{equation}
The friction experienced by the intruder is dominated by the effective 
friction originating from the medium.  

In the ballistic limit, collisions dominate over velocity relaxation. 
Expanding $e^{-\frac{\zeta_{0}}{m}t}$ in Eq.~(\ref{key}) to second order, 
we get 
\begin{equation}
 t_{c}=\sqrt{\frac{2ml_{0}}{F_{ex}}}\ll
 \frac{m}{\zeta_{0}}\,.
\end{equation}
The ballistic limit is given by the presence of pulling forces very large 
compared to the bare friction,
\begin{equation}\label{bc}
\frac{F_{ex}}{\zeta_{0}^{2}}\gg\frac{2l_{0}}{m}.
\end{equation}
The average velocity and the friction of the probe are 
\begin{equation}
\begin{split}
\la v_{s}\ra&=\sqrt{\frac{l_{0}F_{ex}}{2m}}\propto\sqrt{F_{ex}}\\
\zeta&=\sqrt{\frac{2mF_{ex}}{l_{0}}}\propto\sqrt{F_{ex}}\,.
\end{split}
\end{equation}
Both the velocity as well as the friction are proportional to the 
square-root of the external pulling force and independent of the bare 
friction.

\begin{figure}[htb]
\includegraphics[width=\columnwidth]{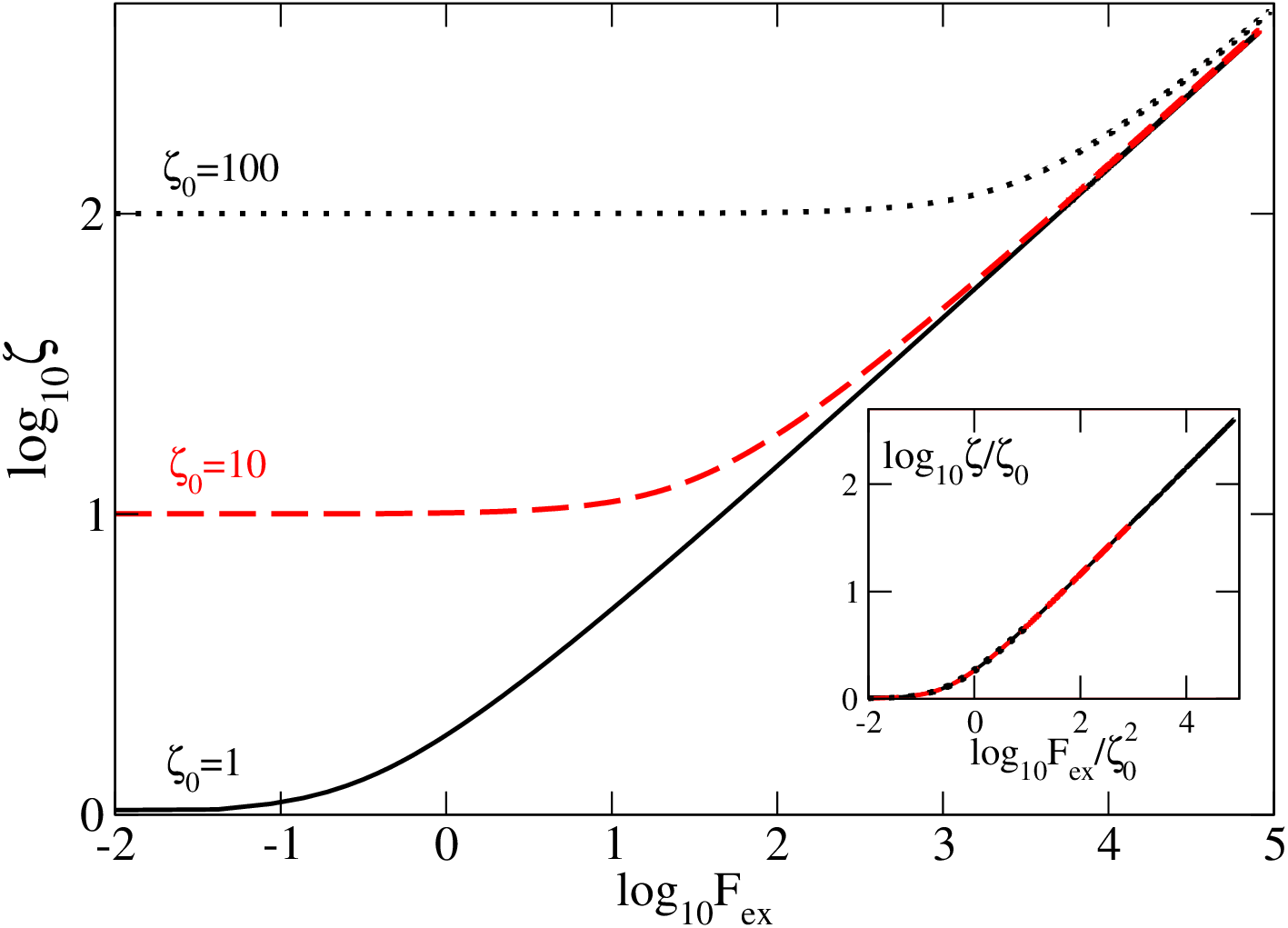}
\caption{\label{fzs} Force-friction relation at large forces for different 
damping in the low-density limit.}
\end{figure} 

The general solution of Eqs.~(\ref{key},\ref{key2}) can be calculated in 
parametric form and is given in Fig.~\ref{fzs}, where the crossover is 
shown for the friction coefficient from a constant (linear large-force 
behavior of the velocity) to the square-root increase (square-root 
increase of the velocity for large forces). The reason why for a driven 
granular system the friction of the probe increases as $\zeta\propto 
\sqrt{F_{ex}}$ in the large-force regime but for colloidal hard-sphere 
systems, the friction only decreases to a constant value can be explained 
as follows. For different bare frictions $\zeta_{0}$, the $F_{ex}$-$\zeta$ 
plots can be rescaled as $F_{ex}/\zeta_{0}^{2}$ versus $\zeta/\zeta_{0}$, 
cf. the inset in Fig.~\ref{fzs}. The behavior of the probe in the 
large-pulling-force regime is determined by the ratio of the collision 
time scale over the Brownian velocity relaxation time scale, 
$t_{c}/\frac{m}{\zeta_{0}}$, or equivalently the value of the rescaled 
force $F_{ex}m/(\zeta_{0}^{2}l_{0})$. In a driven granular system, the 
bare friction is quite small compared with the one in a Brownian 
suspension, $\zeta_{0}=1$ in the granular simulation \cite{Fiege2012} and 
$\zeta_{0}=50$ in the colloidal one \cite{Gnann2011b}. Indeed, one would 
also obtain the same asymptotic behavior $\zeta\propto\sqrt{F_{ex}}$ for 
Brownian systems for extremely large pulling forces.

\section{Conclusion}

We have investigated the microrheology of the driven granular hard sphere 
system by a schematic model and a simple kinetic theory. For small and 
moderate external pulling forces, the schematic model agrees reasonably 
well with the simulation data, cf. Fig.~\ref{fri_fit}, and implies that 
the glass-transition density increases with smaller coefficient of 
restitution $\varepsilon$, confirming predictions from mode-coupling 
theory \cite{Kranz2010,Sperl2012,Kranz2013}. For large forces, glassy 
dynamics becomes irrelevant and a simple kinetic theory clarifies the 
origin of the scaling of the friction with increasing pulling force. When 
damping by a surrounding fluid dominates the motion of the intruder at 
high forces, a second linear emerges where the friction becomes constant. 
When collisions dominate, the friction increases in a square-root law, 
$\zeta\propto\sqrt{F_{ex}}$.

\begin{acknowledgments}
We acknowledge support from DAAD and DFG within FOR1394. We thank Andrea 
Fiege, Matthias Fuchs, Till Kranz, Thomas Voigtmann, Anoosheh Yazdi, and 
Peidong Yu for valuable discussions.
\end{acknowledgments}

\bibliography{pre_ref}
\end{document}